\documentclass[11pt]{article} 
\usepackage{mystyle-new}
\usepackage{epsfig,amsmath} 
\usepackage{hepnames,hepunits}
\usepackage{hyperref}
\usepackage{color}
\usepackage{graphicx}
\definecolor{red}{rgb}{1,0,0}
\def\lesssim{\ \hbox{\raise 2pt \hbox{$<$} \kern -13pt
                     \lower 3pt \hbox{$\sim$}}\ }
\def\greatersim{\ \hbox{\raise 2pt \hbox{$>$} \kern -13pt
                     \lower 3pt \hbox{$\sim$}}\ }

\def\cascade{{\sc Cascade}}
\def\pythia{{\sc Pythia}}
\def\herwig{{\sc Herwig}}

\def\mcatnlo{{MC@NLO}}

\input epsf.tex
\def\desepsf(#1 width #2){\epsfxsize=#2 \epsfbox{#1}}
\def\kt{\ensuremath{k_t}}
\def\ktz{\ensuremath{k_{t,0}}}
\def\pt{\ensuremath{p_{\rm T}}}

\newcommand{\Pmax}{\mu^2}

\newcommand{\alphas}{\ensuremath{\alpha_\mathrm{s}}}

\newcommand{\PBM}{PB}

\newenvironment{tolerant}[1]{\par\tolerance=#1\relax}{ \par }
\usepackage{amsmath,bm}
\usepackage{lineno}

\usepackage{cite,./mcite}
\usepackage{tikz}
\usepackage[symbol]{footmisc}

\newcommand{\dglap}{Gribov:1972ri,Lipatov:1974qm,Altarelli:1977zs,Dokshitzer:1977sg}

\providecommand{\DOI}[1]{\href{http://dx.doi.org/#1}}

\begin{document}

\begin{flushright}
CERN-TH-2019-095\\
DESY 19-087\\ 
\end{flushright}

\begin{center} {\sffamily\Large\bfseries
Production of \boldmath\PZ -bosons 
                           in the parton branching method }
 \\ \vspace{0.5cm}
{ \Large 
A.~Bermudez~Martinez$^{1}$,
P.L.S.~Connor$^{1}$,
D.~Dominguez~Damiani$^{1}$,
L.I.~Estevez~Banos$^{1}$,
F.~Hautmann$^{2,3,4,5}$,
H.~Jung$^{1}$,
J.~Lidrych$^{1}$,
M.~Schmitz$^{1}$,
S.~Taheri~Monfared$^{1}$,
Q.~Wang$^{1,6}$,
R.~\v{Z}leb\v{c}\'{i}k$^{1}$
}\\ \vspace*{0.15cm}
{\large $^1$DESY, Hamburg}\\
{\large $^2$RAL, Chilton OX11 0QX and University of Oxford, OX1 3NP} \\
{\large $^3$Elementary Particle Physics, University of Antwerp, B 2020  Antwerp}\\
{\large $^4$UPV/EHU, University of the Basque Country, E 48080 Bilbao}  \\
{\large $^5$CERN, Theoretical Physics Department, CH 1211 Geneva}\\
{\large $^6$School of Physics, Peking University} \\
\end{center}
\begin{abstract}

Transverse Momentum Dependent  (TMD) parton distributions obtained from 
the Parton Branching (PB) method are combined with next-to-leading-order (NLO) 
calculations of Drell-Yan (DY) production. We apply the \mcatnlo\ method for the hard process calculation and 
matching with the PB TMDs. 
We compute   predictions for   the transverse momentum, rapidity  and $\phi^*$ spectra  of \PZ  bosons.   We find that 
the theoretical uncertainties of the predictions  are dominated by the renormalization and factorization scale dependence,  
  while the impact of  
  TMD uncertainties is moderate.  The theoretical predictions  agree well, within uncertainties,   with  measurements at the Large Hadron Collider (LHC).  
  In particular, we study  the region of lowest  transverse momenta  at the LHC, and comment on  its  sensitivity to  nonperturbative TMD  contributions. 

\end{abstract}

\section{Introduction} 
\label{Intro}
The production of \PZ -bosons in a Drell-Yan (DY) process \cite{Drell:1970wh} in $\Pp\Pp$ collisions is one of the most precisely  measured processes at  high energies at the LHC \cite{Khachatryan:2016nbe,Aad:2015auj,Aad:2014qja,Chatrchyan:2011wt}. At large transverse momenta \pt\ of the \PZ -bosons (for $\pt > {\cal O}(m_{\PZ }))$, higher-order calculations in perturbative QCD including several additional jets are necessary to describe the measurements, while at small transverse momenta ($\pt < {\cal O}(m_{\PZ }))$ soft-gluon perturbative resummations  and nonperturbative contributions are  
needed \cite{Dokshitzer:1978yd,Parisi:1979se,Curci:1979am,Altarelli:1984pt,Collins:1984kg}. An accurate  description of the low-\pt\  spectrum of the vector bosons is   important  for precision measurements of the \PW -boson mass $m_{\PW} $, since it influences the simulation of the transverse mass or the transverse momentum of the decay leptons  from which $m_{\PW} $ is extracted.

Soft-gluon resummation in the low-\pt\  spectrum can be achieved by analytic  resummation 
methods \cite{Bizon:2018foh,Bizon:2019zgf,Catani:2015vma,Scimemi:2017etj,Bacchetta:2019aa,Bacchetta:2018lna,Ladinsky:1993zn,Balazs:1997xd,Landry:2002ix,resbosweb,Alioli:2015toa,Bozzi:2019vnl,Baranov:2014ewa} 
  or by parton shower in multi-purpose Monte Carlo (MC) event generators \cite{Sjostrand:2014zea,Bellm:2015jjp,Bahr:2008pv,Gleisberg:2008ta} 
matched with higher-order matrix elements \cite{Frixione:2003ei,Frixione:2002ik,Frixione:2007vw,Nason:2012pr,Alwall:2014hca,Frederix:2015eii}. Nonperturbative  effects can be incorporated via 
transverse momentum dependent  parton distribution functions  (TMD PDFs, or TMDs) \cite{Angeles-Martinez:2015sea} at low transverse momenta.   

In Refs. \cite{Hautmann:2017xtx,Hautmann:2017fcj} a Parton Branching (PB) approach has been proposed which, similarly to parton shower event generators,  is based 
on the unitarity picture~\cite{Webber:1986mc} of parton evolution, but, unlike these event generators,  uses this picture    to define and evaluate TMD distribution functions. 
The PB method~\cite{Hautmann:2017xtx,Hautmann:2017fcj}  incorporates  soft-gluon angular ordering   
in the parton evolution and   running coupling. This   enables it    to achieve 
leading-logarithmic (LL) and next-to-leading-logarithmic (NLL) accuracy in the soft-gluon resummation, consistently with  
  the formulation~\cite{Marchesini:1987cf,Catani:1990rr} of coherent branching. In this respect, the results of the PB method can be compared with 
results based on the  resummation method~\cite{Collins:1984kg}.  The main feature of the PB approach, compared to parton shower event generators, is 
that TMDs can be obtained and fitted to experimental data, so that the non-perturbative parameters  can be fixed, and predictions can then be constructed with no 
further free parameters. 

In this article we apply  the PB approach to  \PZ -boson   DY production at the LHC.  We use the TMDs obtained within the PB method and 
fitted~\cite{Martinez:2018jxt} to  inclusive deep-inelastic scattering (DIS) precision data from HERA  together with a 
 next-to-leading-order (NLO) calculation of the DY process. All parameters are kept as obtained from the fits to HERA DIS data, and no further adjustment is performed. 
The article is organized as follows.  In Sec.~2 we describe the TMDs obtained from the PB method. In Sec.~3 we describe 
the DY NLO calculation.  In Sec.~4 we present results for the  \PZ -boson   spectra in transverse momentum, rapidity and $\phi^*$ variable. We  compare  our 
results with measurements from the  LHC Run-I at $\sqrt{s} = $ 8 TeV, and present   predictions for  $\sqrt{s} = $  13 TeV.  We give conclusions in Sec.~5. 

\section{Parton distributions from the PB method}
\label{PBTMD}
The PB  method allows evolution equations for  collinear \cite{\dglap } and TMD parton distributions to be solved  numerically in an iterative procedure, by 
making use of the concept of resolvable and non-resolvable branchings and by applying Sudakov form factors to describe the evolution from one scale to another without resolvable branching. The \PBM\ method is described in detail  in Refs. \cite{Martinez:2018jxt,Hautmann:2017fcj,Hautmann:2017xtx}. 

 \begin{tolerant}{1800}
As discussed in detail in \cite{Martinez:2018jxt}, the TMD parton density distributions are obtained from convoluting the perturbative evolution kernel $ {\cal  K}$
with the non-perturbative starting distribution ${\cal A}_{0,b} (x',\ktz^2,\mu_0^2)$:
\end{tolerant}
\begin{eqnarray}
x{\cal A}_a(x,\kt^2,\mu^2) 
 &= &x\int dx' \int dx'' {\cal A}_{0,b} (x',\ktz^2,\mu_0^2) {\cal  K}_{ba}\left(x'',\ktz^2,\kt^2,\mu_0^2,\Pmax\right) 
 \delta(x' x'' - x) 
\nonumber  
\\
& = & \int dx' {\cal A}_{0,b} (x',\ktz^2,\mu_0^2)
\frac{x}{x'} \ { {\cal  K}_{ba}\left(\frac{x}{x'},\ktz^2,\kt^2,\mu_0^2,\Pmax\right) }  \;\; .
\label{TMD_kernel}
\end{eqnarray}
In general, the starting distribution ${\cal A}_0$ at scale $\mu_0$, where $\mu_0 \approx {\cal O} (1 \,  {\rm{GeV}} )$,  
can have flavor-dependent and $x$-dependent $\ktz$ distributions. However,  for  maximal simplicity  we use  
here a factorized form, 
\begin{eqnarray}
{\cal A}_{0,b} (x,\ktz^2,\mu_0^2) & = & f_{0,b} (x,\mu_0^2) \cdot \exp(-| \ktz^2 | / \sigma^2)  \;\; , 
\label{TMD_A0}
\end{eqnarray}
in which the intrinsic $\ktz$ distribution is given by a Gauss distribution  with $ \sigma^2  =  q_0^2 / 2 $  for all parton flavors and all $x$,  with a constant value $q_0 = 0.5$~\GeV. 

Collinear and TMD parton distributions were obtained in \cite{Martinez:2018jxt}  from fits of the parameters of the starting distribution $f_{0,b} (x,\mu_0^2) $ to  the  inclusive-DIS precision 
measurements from HERA,  after  QCD evolution and convolution with the coefficient functions at NLO. Two different sets of parton distributions were obtained: Set~1, which corresponds at collinear level to HERAPDF2.0NLO~\cite{Abramowicz:2015mha},  and Set~2, which  differs by the choice of the scale used in the running coupling $\alphas$, namely, it uses the transverse momentum (instead of the evolution scale), corresponding to the angular-ordering approach. An additional parameter $q_{cut}=1~\GeV$ is introduced in $\alphas(\max(q^2_{cut},|{\bf q}_{t,}^2|))$ and a model dependence was estimated in Ref.~ \cite{Martinez:2018jxt} with a variation around the default choice.

\begin{figure}[h!tb]
\begin{center} 
\includegraphics[width=0.48\textwidth]{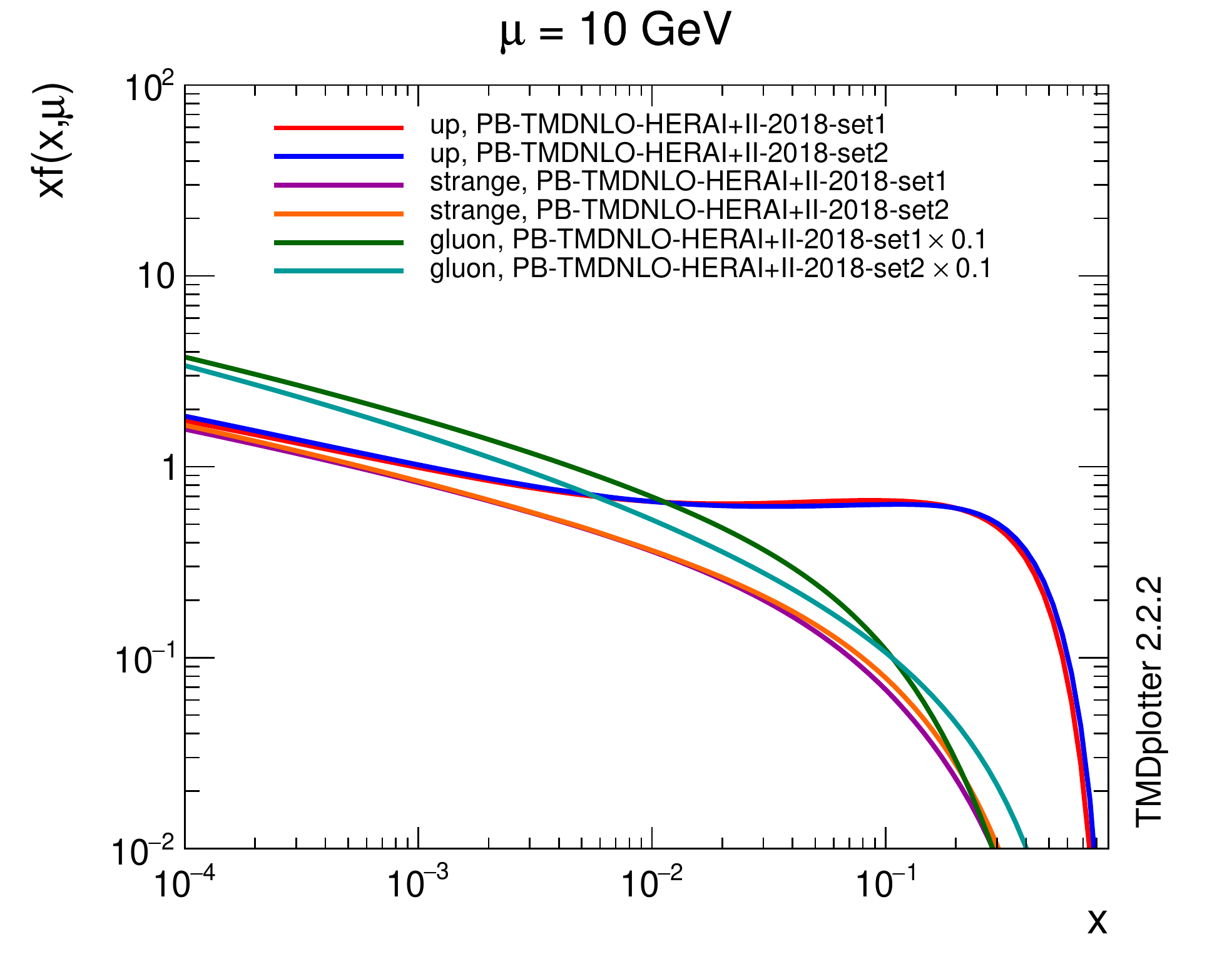} 
\includegraphics[width=0.48\textwidth]{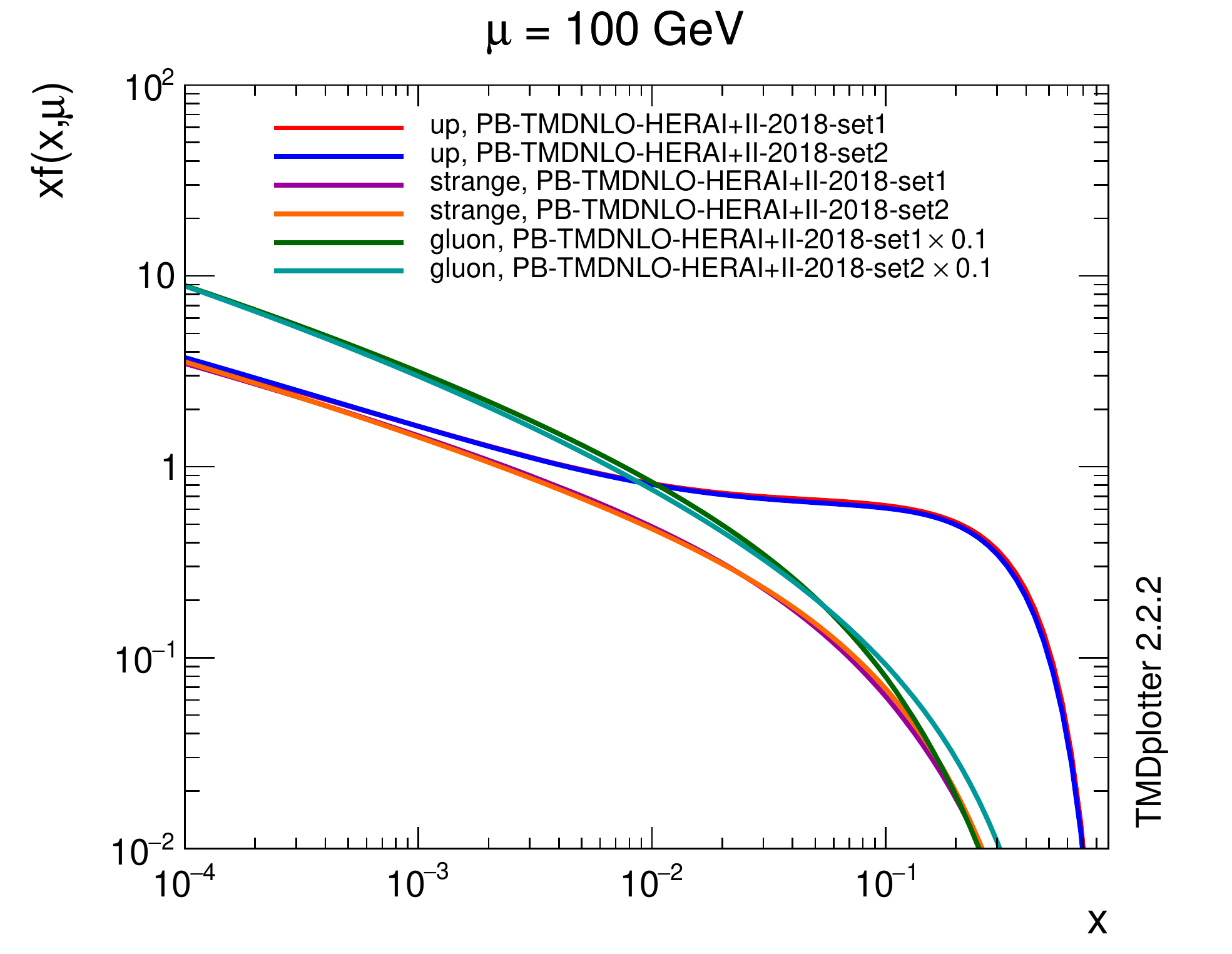} 
  \caption{\small Collinear parton distributions for up, strange and gluon  (PB-NLO-2018-Set1 and PB-NLO-2018-Set 2) as a function of $x$ for different scales $\mu$.
  }
\label{TMD_pdfs1}
\end{center}
\end{figure}

In Fig.~\ref{TMD_pdfs1} the collinear parton densities are shown for up-quark, strange-quark  and gluon distributions at evolution scales of $\mu=10, 100 $ \GeV . At $\mu=10$ \GeV\  differences especially for the gluon are observed between Set~1 and Set~2. At scales relevant for \PZ\ production ($\mu=100$ \GeV) the differences between the two sets are small. The collinear parton densities are available in a format compatible with the one employed in LHAPDF~\cite{Buckley:2014ana},  and can be used in the calculation of NLO processes.

\begin{figure}[h!tb]
\begin{center} 
\includegraphics[width=0.325\textwidth]{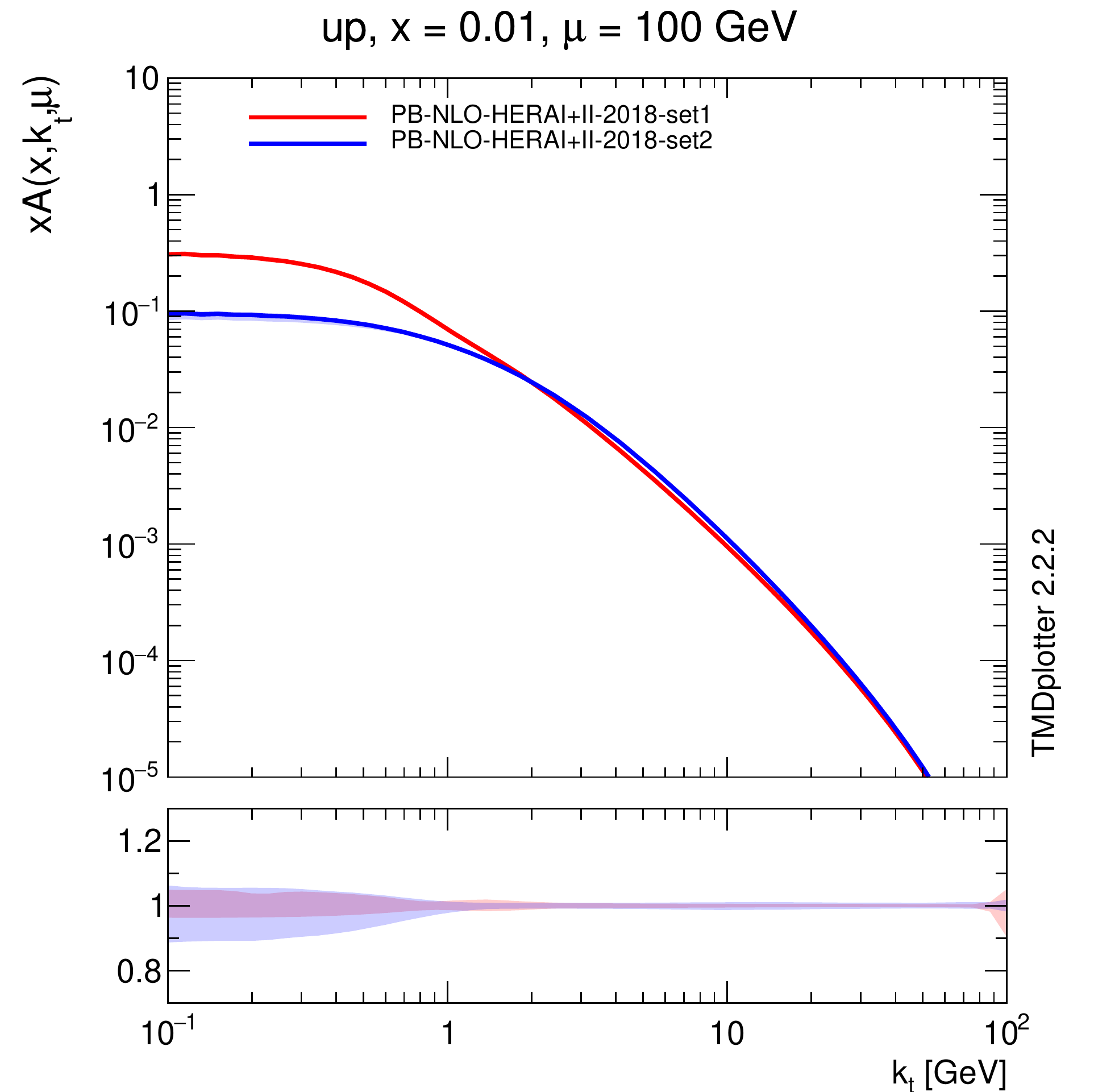} 
\includegraphics[width=0.325\textwidth]{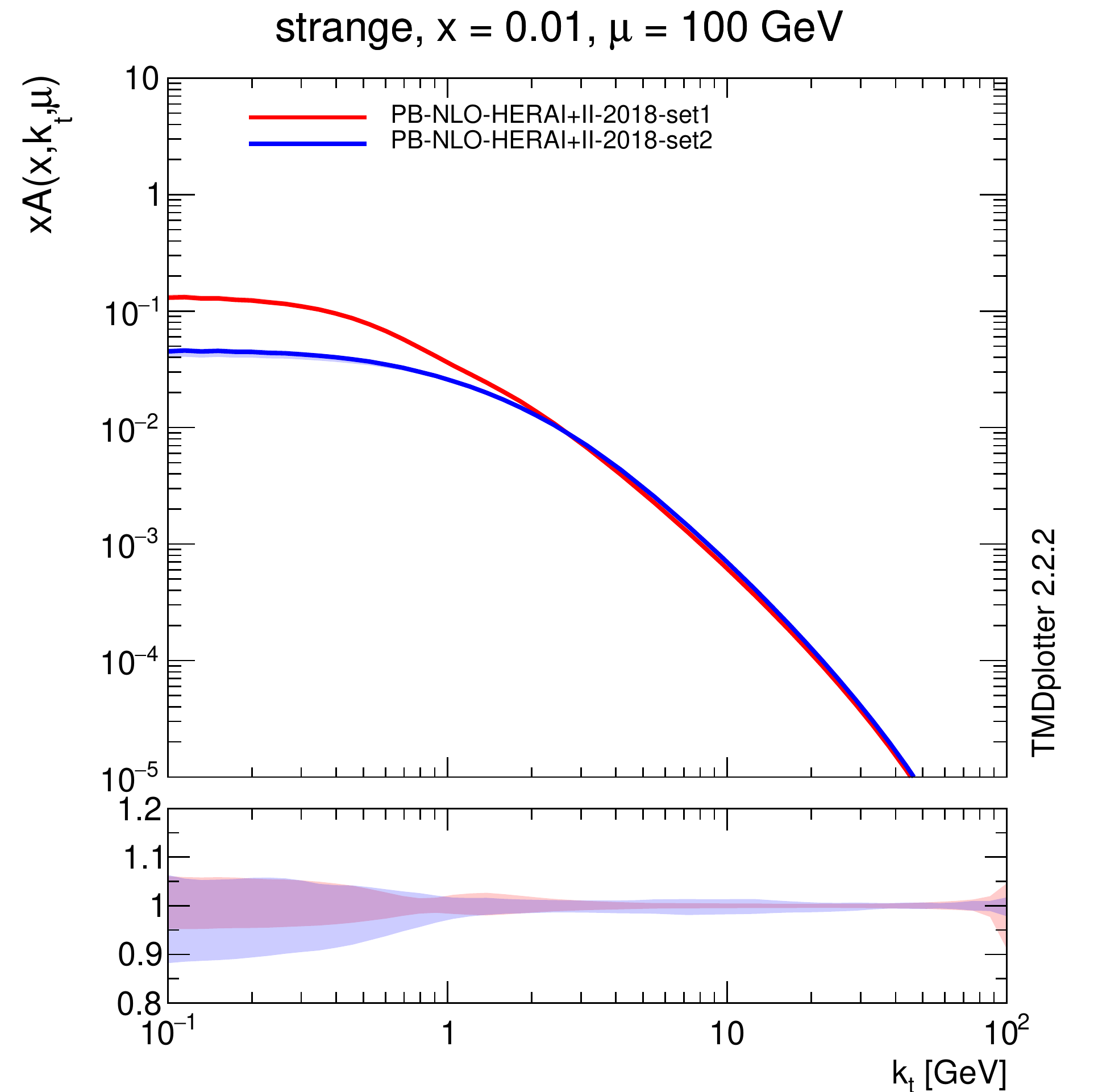} 
\includegraphics[width=0.325\textwidth]{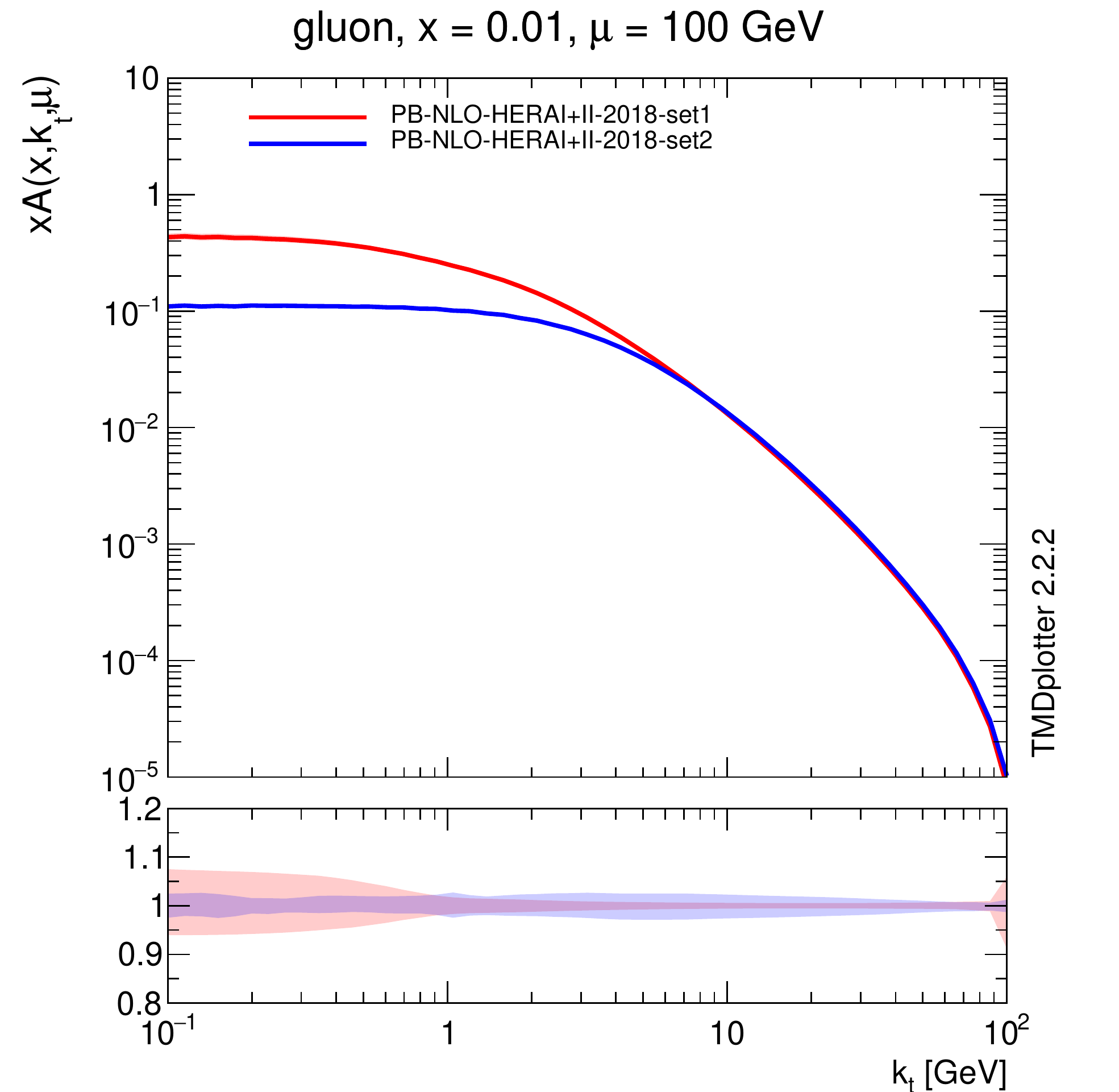} 
  \caption{\small TMD parton distributions for up, strange and gluon  (PB-NLO-2018-Set1 and PB-NLO-2018-Set 2) as a function of $\kt$ at $\mu=100$ \GeV and $x=0.01$. The lower panels show the experimental and model uncertainties with respect to the central values.  }
\label{TMD_pdfs2}
\end{center}
\end{figure}

In Fig.~\ref{TMD_pdfs2}   we show the transverse momentum distributions for up-quark, strange-quark and gluon partons at $x=0.01$ (typical for \PZ\ production at  $\sqrt{s} = 8 $ \TeV ) and  $\mu = 100$ \GeV . The lower panels show the uncertainty bands obtained from experimental and model uncertainties.

\begin{figure}[h!tb]
\begin{center} 
\includegraphics[width=0.45\textwidth]{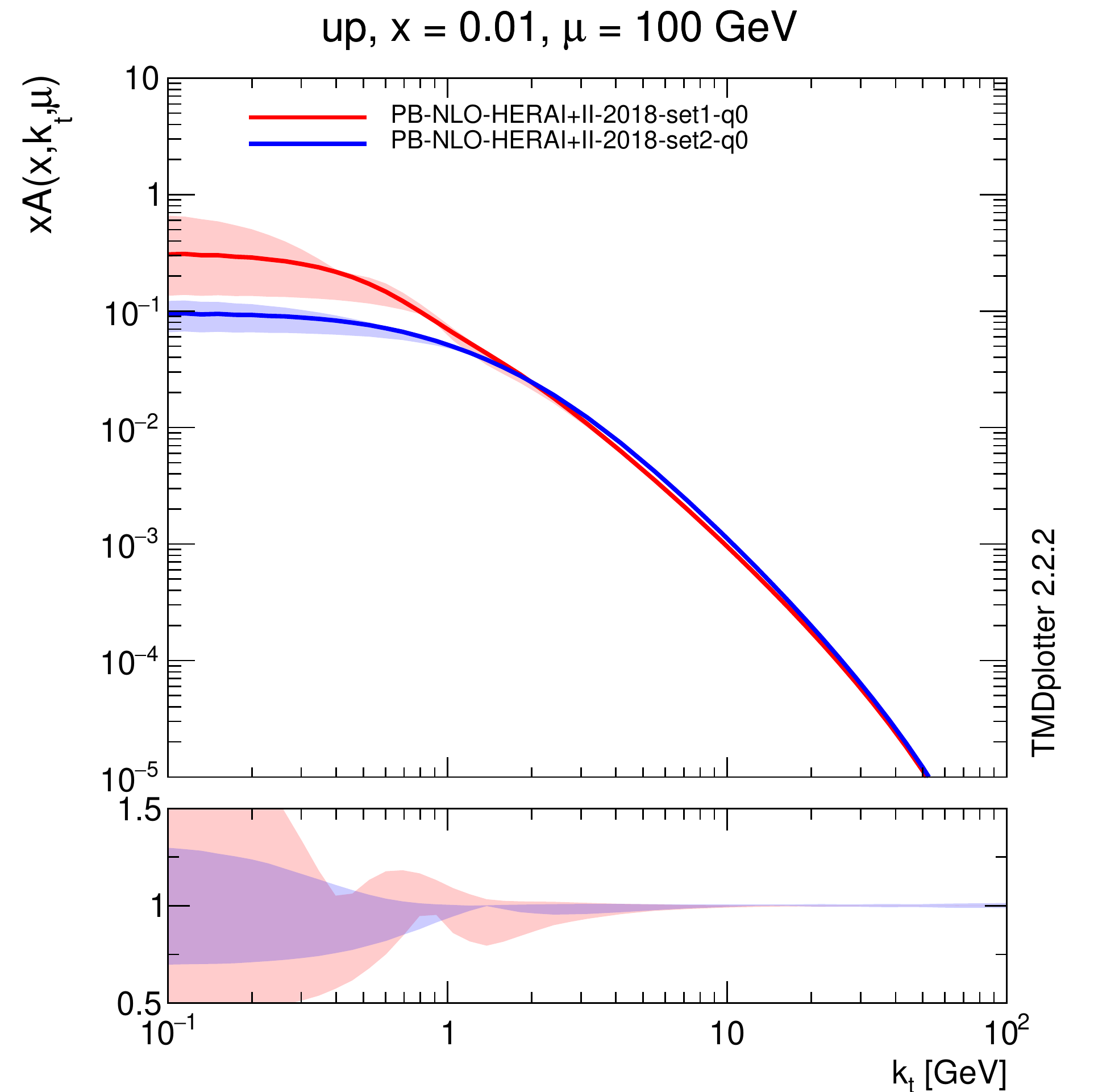} 
\includegraphics[width=0.45\textwidth]{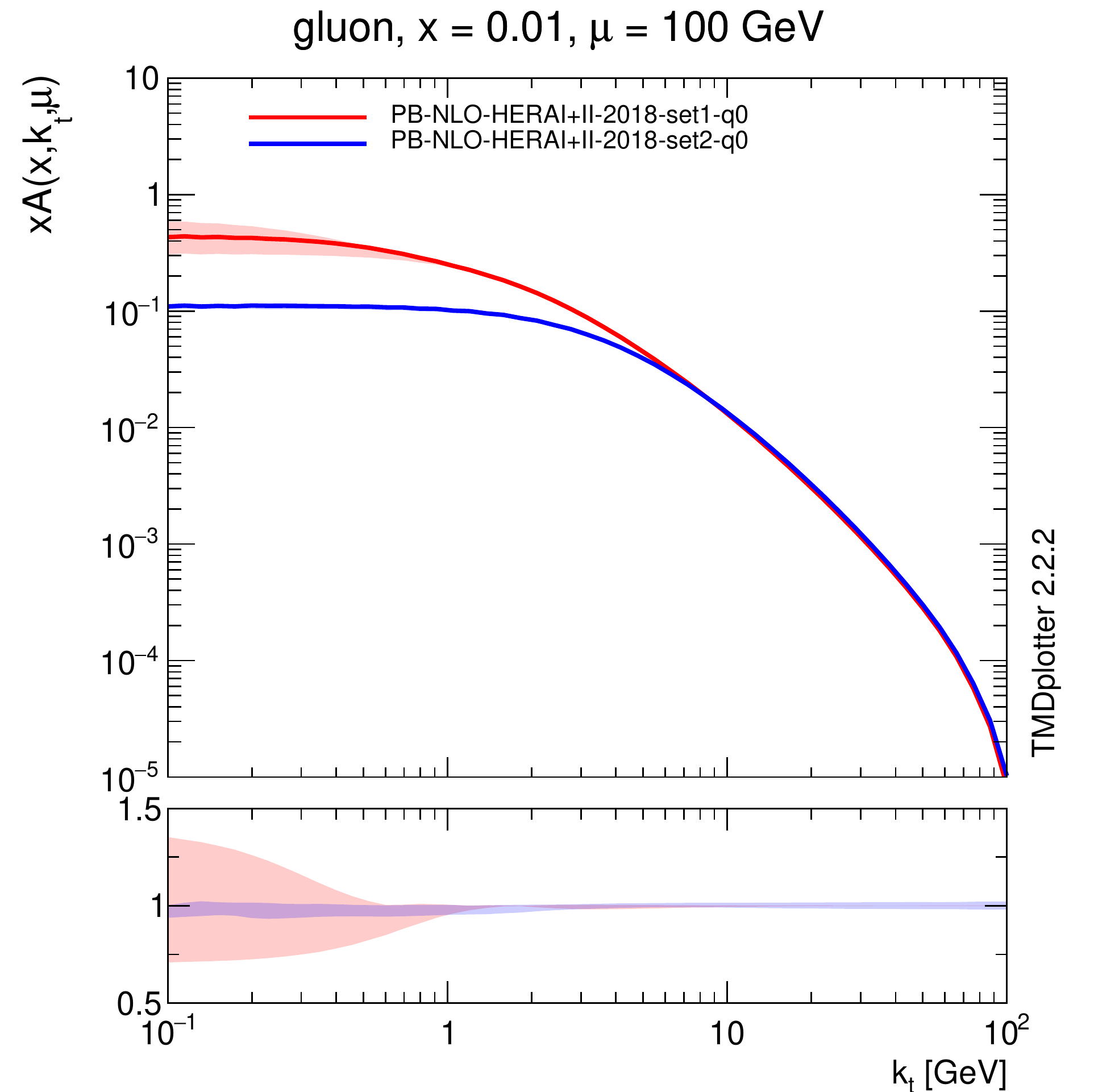} 
  \caption{\small TMD parton distributions for up-quark and gluon  (PB-NLO-2018-Set1 and PB-NLO-2018-Set 2) as a function of $\kt$ at $\mu=100$ \GeV and $x=0.01$. The  band shows the uncertainty coming from a  variation of the mean of the intrinsic \kt\ distribution.
  }
\label{TMD_pdfs3}
\end{center}
\end{figure}

In Fig.~\ref{TMD_pdfs3}  we show the transverse momentum distributions for up-quarks and gluons at  $x=0.01$ and  $\mu = 100$ \GeV . The  band shows the uncertainty coming from changing the width of the Gauss distribution $q_0$ in Eq.~(\ref{TMD_A0}) by a factor of 2 up and down in the fit as described in Ref.~\cite{Martinez:2018jxt}. This  variation will be used later to estimate the uncertainty of the \PZ -boson \pt -spectrum coming from the intrinsic \kt -distribution.

We conclude this section with some comments on Eqs.~(\ref{TMD_kernel}),(\ref{TMD_A0}).  The structure of the TMD distribution is not the same for quarks and gluons in the PB method: both the evolution kernels $ {\cal  K}$ and the intrinsic distributions ${\cal A}_0$ in Eq.~(\ref{TMD_kernel})   are in general different for different parton species.  
Taking  a simple flavor-independent (and $x$-independent) Gaussian in Eq.~(\ref{TMD_A0}) is not a feature of the PB method, but rather it is motivated by our finding in Ref.~\cite{Martinez:2018jxt}   that the precision DIS data from HERA used for the fits are not sensitive to the flavor structure of the intrinsic distribution. 

A similar remark applies to nonperturbative contributions to evolution kernels. At present, the kernel 
$ {\cal  K}$  in Eq.~(\ref{TMD_kernel})   does not include any nonperturbative components. But in principle these could be introduced in the PB framework 
 as nonperturbative contributions to the Sudakov form factors, and parameterized in terms of nonperturbative functions to be determined from fits to experimental data. Similarly 
 to the comment above, this is not done at the moment mainly because HERA and LHC have little sensitivity to these long-distance effects. 

\section{ \PBM -TMDs  and NLO calculation of  Drell-Yan production}
In Ref. \cite{Martinez:2018jxt}  leading order (LO) matrix elements for the calculation of \PZ\ production in pp collisions at the LHC are used. A rather good description was obtained for the small-\pt\  region.  
 Predictions applying  \PBM -TMDs   for the calculation of \PZ -boson production in proton-lead collisions have been recently reported in Ref.~\cite{Blanco:2019qbm}.
In the following we will describe how to use the \PBM\ TMD-parton distributions together with higher order calculations. 
We make use of 
{\sc MadGraph5\_aMC@NLO} (version 2.6.4, in the following labelled \mcatnlo )  \cite{Alwall:2014hca} framework and apply the NLO 
\PBM\ parton distributions with $\alphas(M_{\PZ } )= 0.118$ for the NLO calculations of inclusive Drell-Yan production.

In \mcatnlo\ subtraction terms, corresponding to the parton shower used, are calculated and subtracted from the NLO cross section. Since the \PBM -method follows angular ordering, with the same choice of the argument in \alphas\ and in the kinematics as used in the parton shower generator  \herwig ++\cite{Bahr:2008pv} and \herwig 6 \cite{Corcella:2002jc,Marchesini:1991ch},  the calculation of the hard process with the subtraction terms of \herwig\  is used. NLO splitting functions in the \PBM -method are used (consistent with NLO collinear parton densities in the MC@NLO calculation), while the subtraction terms include only LO splitting functions. However, these differences appear at an order, beyond the present next-to-leading order.

The NLO event generator \mcatnlo\ 
generates events (with weights) which are stored in a format which can be read by parton shower event generators (LHE format) \cite{Alwall:2006yp}. Instead of a parton shower event generator, we apply the \PBM -TMDs and modify the kinematics of the initial state partons (and as a consequence the final state partons and particles) according to the transverse momentum distributions given by the \PBM -TMDs. Since adding transverse momenta requires changes also in the longitudinal momenta, we require that the invariant mass of the partonic system $\hat{s}$ as well as the rapidity of the system is not changed (a method which was applied in \herwig 6 and \herwig ++ \cite{Bahr:2008pv}).
 
\begin{figure}[h!tb]
\begin{center} 
\includegraphics[width=0.47\textwidth]{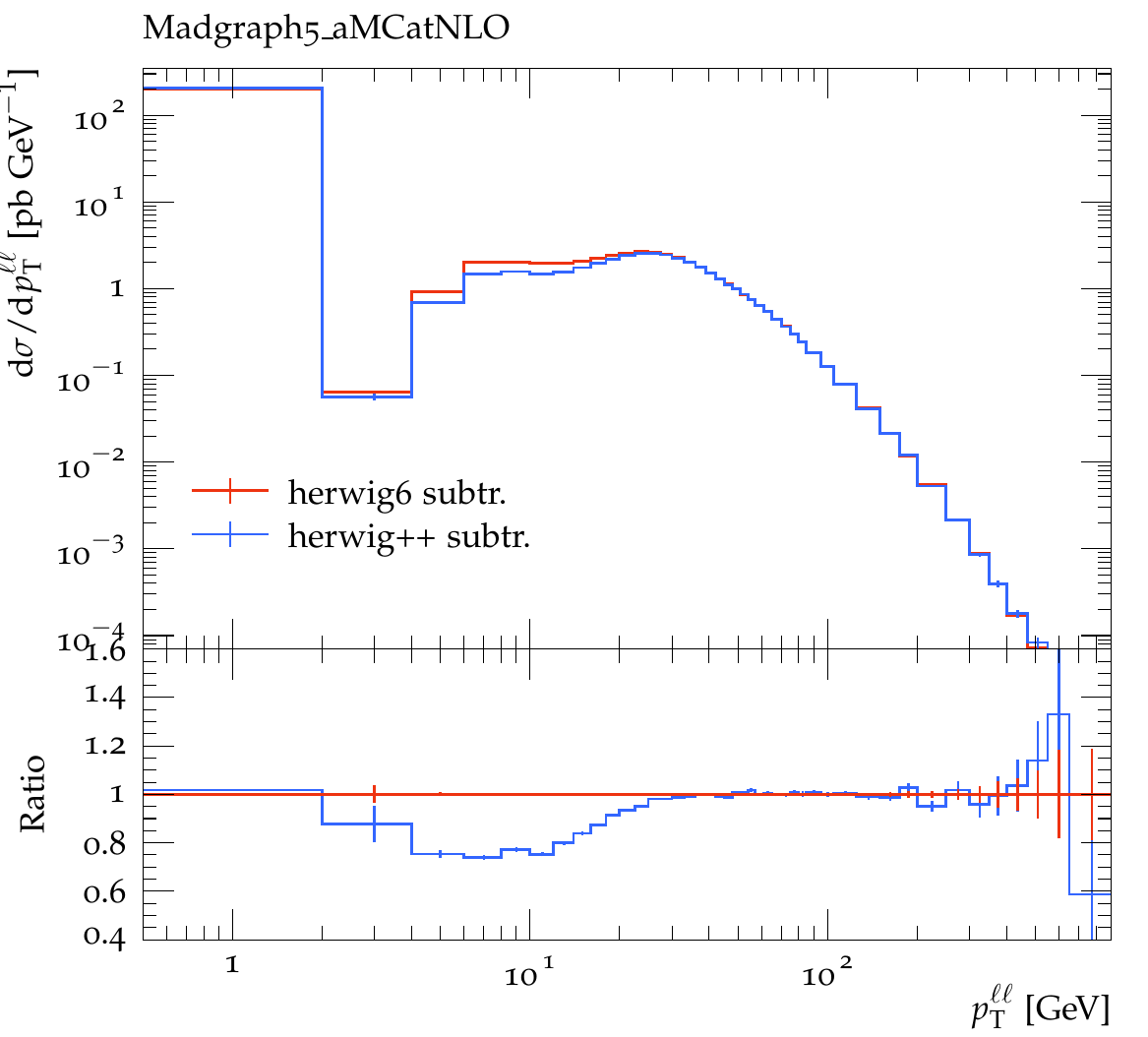} 
\includegraphics[width=0.47\textwidth]{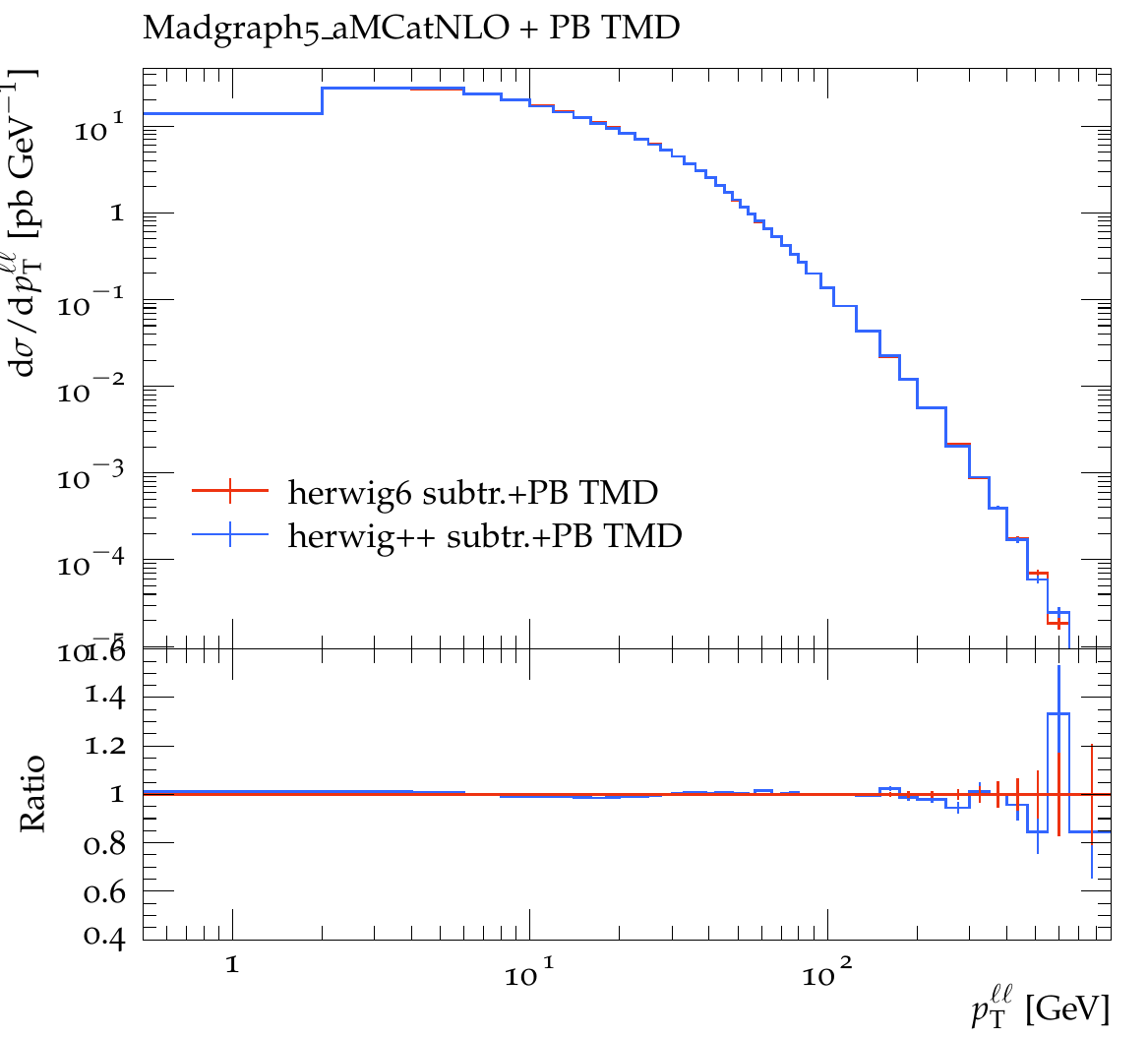} 
  \caption{\small Transverse momentum spectrum of \PZ\  obtained  for different subtraction terms: (left) at parton level (LHE level), (right) after inclusion of \PBM -TMDs.
 }
\label{mcatnlo_lhe}
\end{center}
\end{figure}

 \begin{tolerant}{800}
The transverse momentum spectrum of \PZ -bosons obtained from the calculation of \mcatnlo\ at a purely partonic level (LHE level) is shown in Fig.~\ref{mcatnlo_lhe} (left).
Shown are the distributions obtained with \herwig 6 and \herwig ++ subtraction terms.  In Fig.~\ref{mcatnlo_lhe} (right) the distribution is shown after transverse momenta are added according to the  \PBM -TMDs. The differences between using calculations with the different subtractions terms are very small, as seen in the ratios in the lower panels. The \PBM -TMD contribute to the \pt\ spectrum of the \PZ -boson up to the scale of the hard process, not only in the non-perturbative region (since the TMDs extend to large \kt , as can be already seen from Fig. \ref{TMD_pdfs2}, \ref{TMD_pdfs3}).
\end{tolerant}

In order to avoid double counting between the contribution of the real emission treated by the matrix element calculation and the contribution from the \PBM -TMD (or parton shower), a matching scale $\mu_{m}$ needs to be defined. This scale is determined by the NLO calculation and is transferred to the user via the parameter \verb+SCALUP+ (included  in the LHE file).

The \PBM -TMDs depend, as indicated in 
Eq.~(\ref{TMD_kernel}), on the longitudinal momentum fraction $x$, the factorization scale $\mu$ and the transverse momentum $\kt$. The factorization scale $\mu$ and the longitudinal momentum fraction $x$ are used in the calculation of the hard process for collinear kinematics and 
a scale corresponding to $\mu = \frac{1}{2} \sum_i \sqrt{m^2_i +p^2 _{t,i}}$ is chosen by default, with the sum running over all final state particles, that is, in the case of Z production,  over the decay products and the final jet. 

In order to allow  the full phase space to be covered for the transverse momentum in the \PBM -TMD, the factorisation scale $\mu$ is set to the invariant mass of the hard process $\mu=\sqrt{\hat{s}}$ for the underlying Born configuration. For the real emission configuration the scale is changed  to $\mu=\frac{1}{2} \sum_i \sqrt{m^2_i +p^2 _{t,i}}$, as in the \mcatnlo\ calculation. Another possible choice is to set $\mu$
to the maximum transverse momentum of the most forward and backward parton to ensure a proper matching with the angular ordering evolution of the \PBM -TMDs; the results are very close to the results obtained with $\mu=\frac{1}{2} \sum_i \sqrt{m^2_i +p^2 _{t,i}}$.
Finally, the transverse momentum is constrained to be smaller than the matching scale $\mu_{m}=$\verb+SCALUP+. The calculation are performed with the \cascade 3 package ~\cite{Jung:2010si,Jung:2001hx} (version  \verb+3.0.X+), which allows to read LHE files and to produce output files to be analysed with Rivet~\cite{Buckley:2010ar}.

The resulting transverse momentum spectrum of \PZ -bosons obtained from the calculation of \mcatnlo\ after inclusion of \PBM -TMDs for different subtraction terms is shown in Fig.~\ref{mcatnlo_lhe} (right). It is interesting to note that,  for 
the \pt\ spectrum of \PZ -bosons calculated at inclusive level, no sensitivity to the subtraction terms is observed after the \PBM -TMDs are included, because the difference in 
Fig.~\ref{mcatnlo_lhe} (right) translates into an effect of at most a percent in the total rate in Fig.~\ref{mcatnlo_lhe} (left). It has been checked explicitly that a similar result is obtained when using \pythia 8 \cite{Sjostrand:2014zea} parton showers.

\subsection{\boldmath \PBM -method and parton showers}

The basic principles of the \PBM -method and parton showers are very similar: both methods rely on the definition of resolvable branchings and treat every branching separately, where the kinematics can be reconstructed. 

The \PBM -method is a method to determine the parton density, and has been applied to precision DIS measurements from HERA to determine the free parameters of the starting distributions \cite{Martinez:2018jxt}. The parton densities  (collinear as well as transverse momentum dependent ones) have been obtained with NLO DGLAP splitting functions and two-loop \alphas\ with $\alphas=0.118$. The evolution scale and the definition of resolvable branching is fixed.
The intrinsic \kt -distribution is not really constrained from inclusive DIS measurements, and a variation of the mean of the intrinsic Gauss distribution leads to the same fits of collinear parton densities. An uncertainty is assigned coming from a variation of the mean of the Gauss distribution by a factor of two, as well as uncertainties coming from the experimental uncertainties of the data points and model uncertainties as discussed in Ref.~\cite{Martinez:2018jxt}.  
Apart from the scale uncertainties of the hard process calculation and uncertainties of the matching scale, all parameters are fixed.

In a traditional parton shower approach (e.g. \cite{Sjostrand:2014zea,Bellm:2015jjp,Bahr:2008pv,Gleisberg:2008ta}) the evolution is not constrained by the parton density,  and  \pt - ordered or angular ordered evolution are used with any collinear parton density, together with specific choices on the definition of resolvable branchings. The free parameters are usually constrained by fits to measurements, the Monte Carlo (MC) event generator tunes, provided by the MC authors directly or by experiments (e.g. \cite{Sirunyan:2019dfx,Khachatryan:2015pea,ATL-PHYS-PUB-2014-021,ATL-PHYS-PUB-2013-017}). The value of \alphas\ is usually different from the one used in the parton density, as discussed for example in Ref.~\cite{Frederix:2015eii},  and special tunes are needed for LO or NLO calculations.

While the principles of \PBM -method and parton shower are similar, the major advantage of the \PBM -TMDs lies in a consistent use of parton evolution, the concept of resolvable branching and the determination of the transverse momentum during the evolution in a way which is constrained by fits to inclusive data, as the usual collinear parton densities. The central values of the predictions might be very similar in both approaches, however the uncertainties obtained in the \PBM -TMD come directly from the fit to inclusive DIS data, while the ones coming from a traditional parton shower are obtained separately and are not constrained by fits to parton densities. Numerically the  uncertainties for the \pt -spectrum of the \PZ -boson  in the \PBM -method are of the order of a percent or smaller (see discussion around Fig.~\ref{Zpt-TMD_uncertainty}),  while in a parton shower approach the uncertainties range from 50\% at LO to ca. 5\% for NLO splitting functions \cite{Hoche:2017hno}.

\section{\boldmath\PZ -boson    production at the LHC}
\subsection{Transverse momentum and   $\phi^*$  spectra    at 8 \TeV}

\begin{figure}[htb]
\begin{center} 
\includegraphics[width=0.495\textwidth]{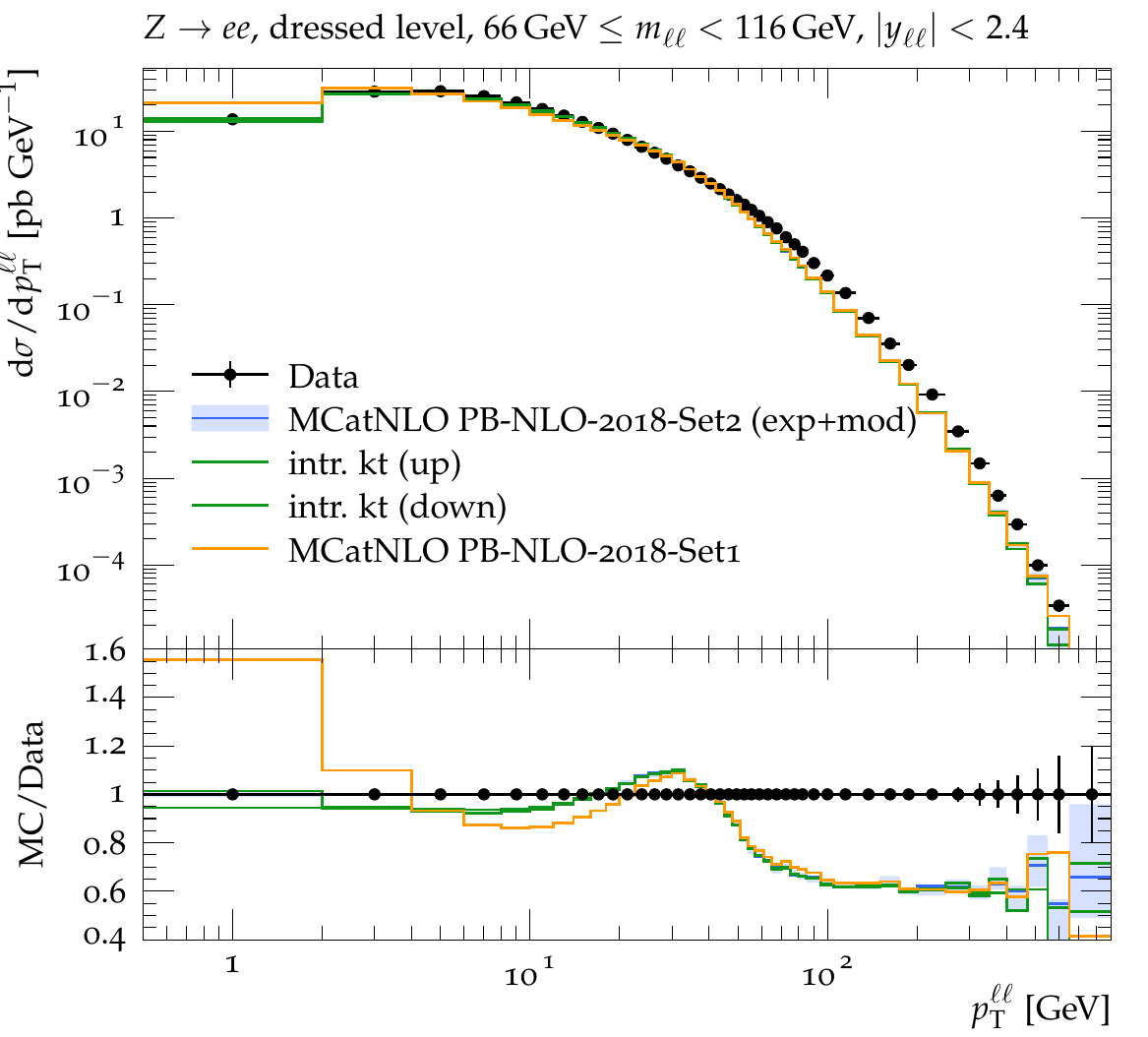} 
\includegraphics[width=0.495\textwidth]{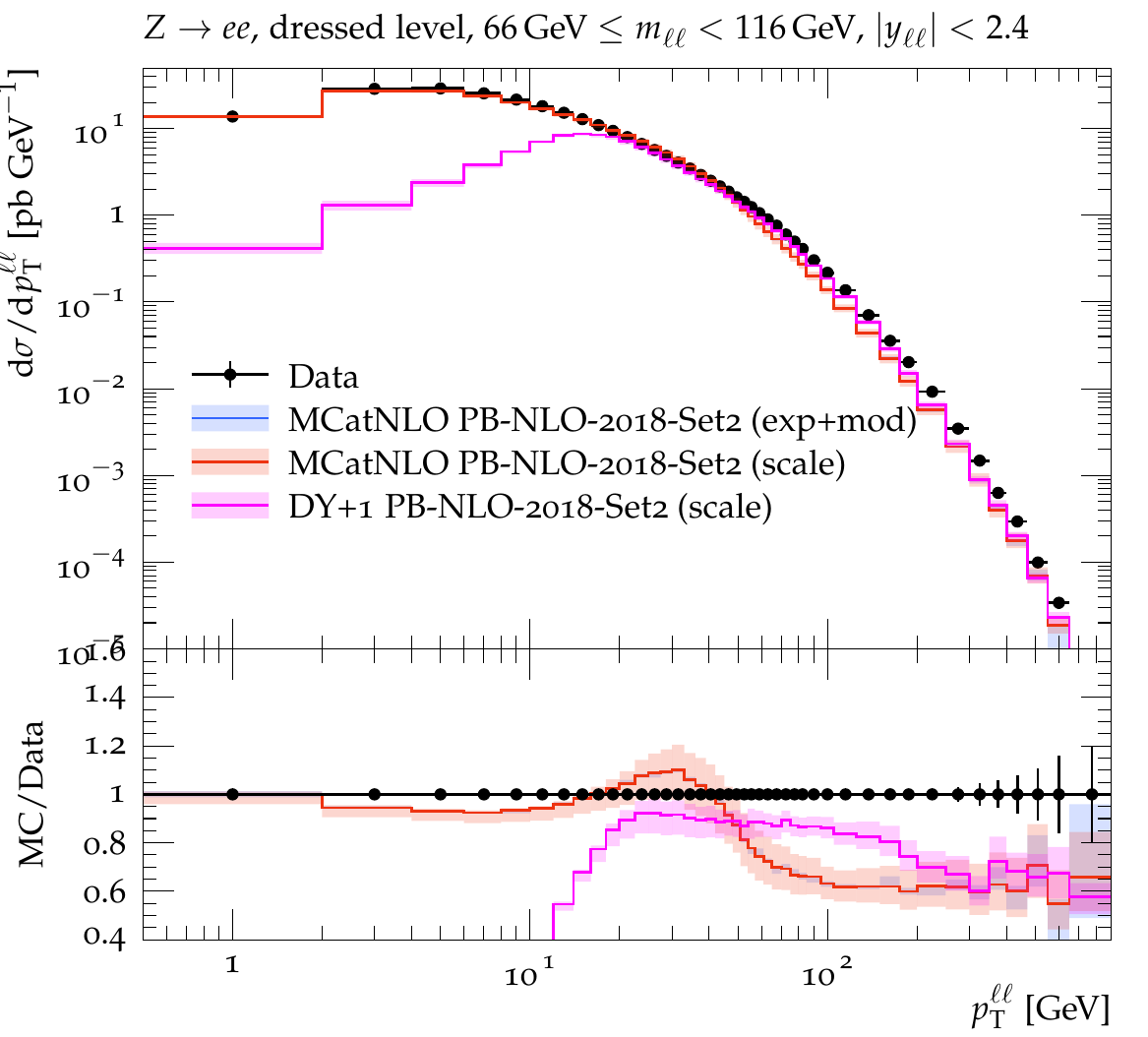} 
\caption{\small Transverse momentum $p_T$ spectrum of \PZ -bosons as measured by \protect\cite{Aad:2015auj} at $\sqrt{s}=8~\TeV$ compared to the prediction from \protect\mcatnlo\ with \protect\PBM -TMD NLO 2018 \protect\cite{Martinez:2018jxt}.
Left: uncertainties from the \protect\PBM -TMD and 
uncertainties coming from changing the width of the intrinsic gauss distribution by a factor of two. Right: with uncertainties from the TMDs  and scale variation combined. 
  }
\label{Zpt-TMD_uncertainty} 
\end{center}
\end{figure}

In Fig.~\ref{Zpt-TMD_uncertainty} we show the prediction for  the transverse momentum spectrum of \PZ -bosons at $\sqrt{s}=8~\TeV$ obtained with a calculation using \mcatnlo\ together with the \PBM -TMD~\cite{Martinez:2018jxt}, and compare it with the measurement by   ATLAS \cite{Aad:2015auj}. The calculation is performed with \cascade ~\cite{Jung:2010si,Jung:2001hx} (version  \verb+3.0.X+)  and Rivet~\cite{Buckley:2010ar}.  Fig.~\ref{Zpt-TMD_uncertainty} (left) shows the prediction using  PB-2018-Set1 and Set2 parton distributions for both the collinear and the TMD calculations. The prediction based on PB-2018-Set1 overshoots the measurement at small transverse momentum of the \PZ -boson, while the prediction based on PB-2018-Set2 agrees very well with the measurement. This difference in the \pt -spectrum is a direct consequence of the differences in the \kt -spectrum of the TMDs, shown in Fig.~\ref{TMD_pdfs3}: in Set~2 the value of $\alpha_s$  at small \kt\  is larger, leading to a higher probability for radiation and therefore a depletion of the distribution at small \kt.  In the following we will use PB-2018-Set2 only. 
In Fig.~\ref{Zpt-TMD_uncertainty} (left) the uncertainties of the TMD PDF are shown which come from experimental and model uncertainties (as in a fit to collinear parton densities), they are very small ($\sim 2 \%$). In addition we show 
the effect of varying the mean of intrinsic \kt -distribution by a factor of two in Fig.~\ref{Zpt-TMD_uncertainty} (left). Only in the lowest \pt\ bin an effect of ca. 5\% can be observed.
The uncertainties coming from the TMD densities  and the uncertainties coming from the scale variation ($\mu_F$ and $\mu_r$ are varied by a factor of 2 up and down independently) in \mcatnlo\ are shown in Fig.~\ref{Zpt-TMD_uncertainty} (right). The uncertainties from the TMD determination are small compared to the uncertainties coming from scale variation.

The bump in the \pt\ distribution at $\sim30~\GeV$ is an effect of the scale choice in \mcatnlo , and it has been explicitly verified  that a similar structure is observed when using \herwig 6 or \pythia\ instead of the \PBM -TMD. The deviation of the prediction at higher transverse momenta comes from missing higher order contributions in the matrix element calculations, as only ${\cal O}(\alphas)$ corrections are included, and the restriction of transverse momentum of the TMD  by $\mu_{m}$.

In Fig.~\ref{Zpt-TMD_uncertainty} (right) the contribution from DY+1 jets (with $\pt^{jet} > 10 ~\GeV$) at NLO obtained with PB-2018~Set 2 parton distribution is shown in addition, with the band showing  the uncertainty coming from the scale variation. At larger \pt\ the higher order contribution plays an important role and improves the description of the measurements. At this stage, we do not attempt to merge DY and DY+1 jet calculations. This will require additional studies  in the use of the TMDs.

\begin{table}[htb]
\renewcommand*{\arraystretch}{1.0}
\centerline{
\begin{tabular}{ c|c|c|c }
$p_T^{cut} $ [\GeV] & $n$& {Set1 } & {Set2 }\\
\hline
20    & 10  & 10.5 & 1.7 \\
30    & 14  & 7.7 & 1.5\\
40     & 17  & 6.3   & 1.3 \\
50    &  21  & 5.2   & 1.2 \\
60    &  24  & 4.8   & 1.8
\end{tabular}}
\caption{\small Values of $\chi^2/n$ as a function of the upper limit ot the \PZ\ transverse momentum $p_T^{cut} $.} 
\label{chi2_tab}
\end{table}

In Table~\ref{chi2_tab} we show a quantitative comparison of our predictions with the measured \pt\ distribution. We calculate $\chi^2/n$ between the $n$ measurement points and the prediction taking into account the experimental uncertainties (statistical, systematic and luminosity uncertainty, added in quadrature) and the theoretical uncertainties (uncertainties from the TMD determination and scale uncertainties, as shown in Fig.~\ref{Zpt-TMD_uncertainty}, added in quadrature). The agreement between the measurement and the prediction using Set~2 is very good for $p_T^{cut}< 50~\GeV$, although no parameters are fitted. The better $\chi^2$ obtained with Set~2 compared to Set~1 supports the use of 
transverse momentum (instead of the evolution scale), corresponding to the angular-ordering approach, as the argument of $\alpha_s$.

\begin{figure}[htb]
\begin{center} 
\includegraphics[width=0.495\textwidth]{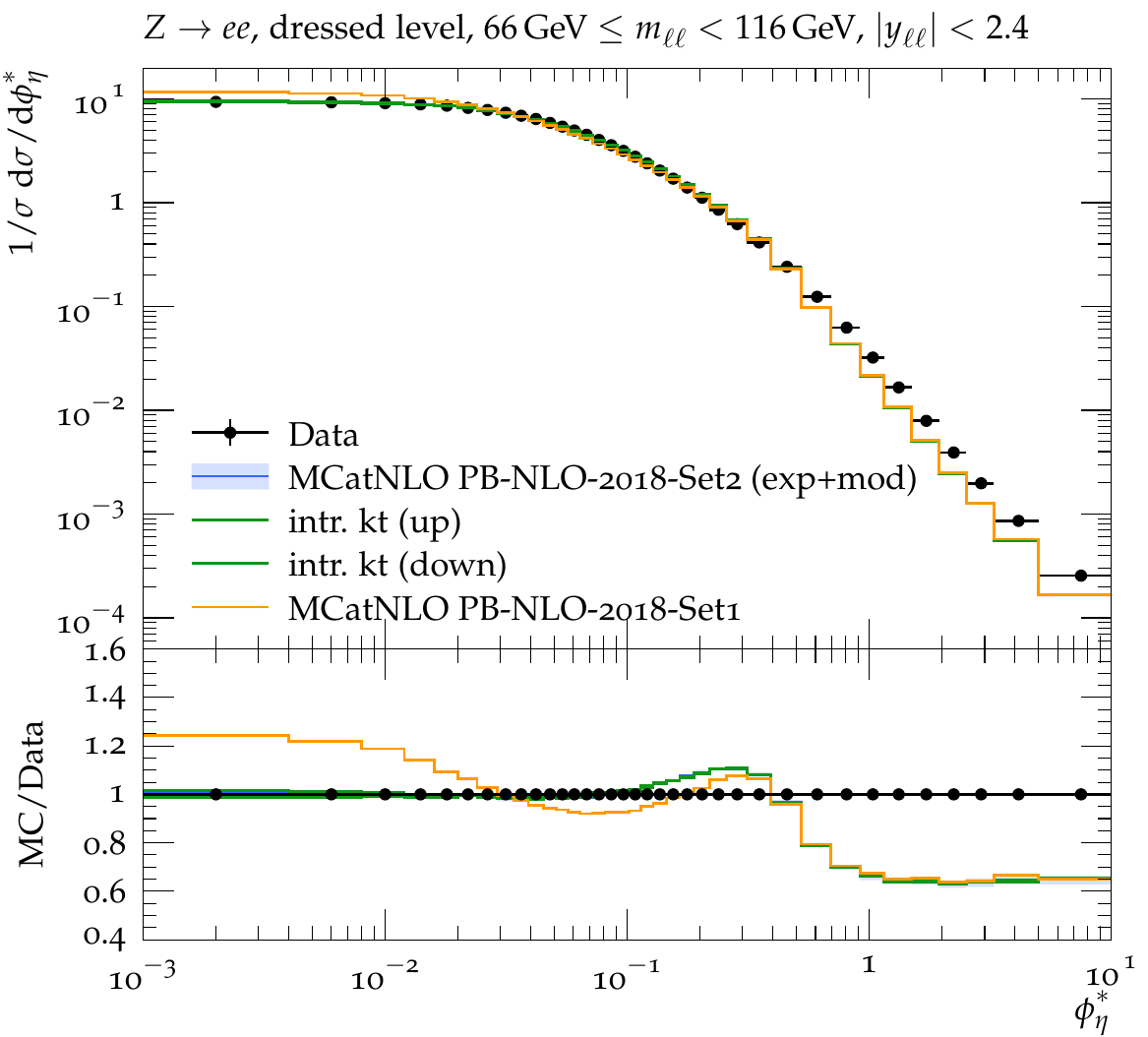} 
\includegraphics[width=0.495\textwidth]{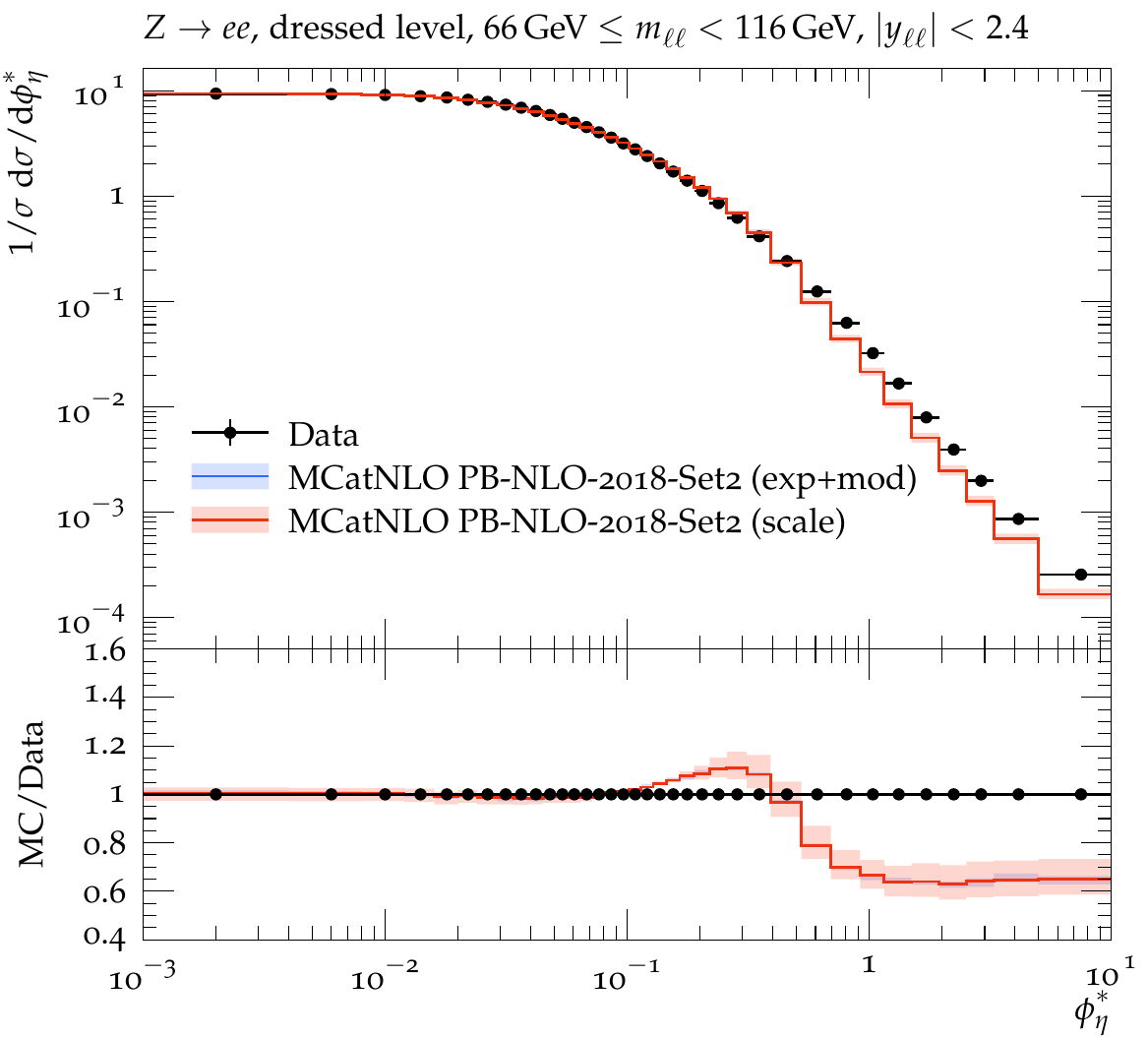} 
\caption{\small  $\phi^*$ spectrum of \PZ -bosons as measured by \protect\cite{Aad:2015auj} at $\sqrt{s}=8~\TeV$ compared to the prediction from \protect\mcatnlo\ with \protect\PBM -TMD NLO 2018 \protect\cite{Martinez:2018jxt}.
Left: uncertainties from the \protect\PBM -TMD and 
uncertainties coming from changing the width of the intrinsic gauss distribution by a factor of two. Right: with uncertainties from the TMDs  and scale variation combined. 
  }
\label{Zphi-TMD_uncertainty} 
\end{center}
\end{figure}

In Fig.~\ref{Zphi-TMD_uncertainty} we show a comparison of the calculation with the $\phi^*$ distribution as measured in \cite{Aad:2015auj}.

\subsection{Predictions for \boldmath\PZ -boson production at 13 \TeV }
\begin{figure}[htb]
\begin{center} 
\includegraphics[width=0.495\textwidth]{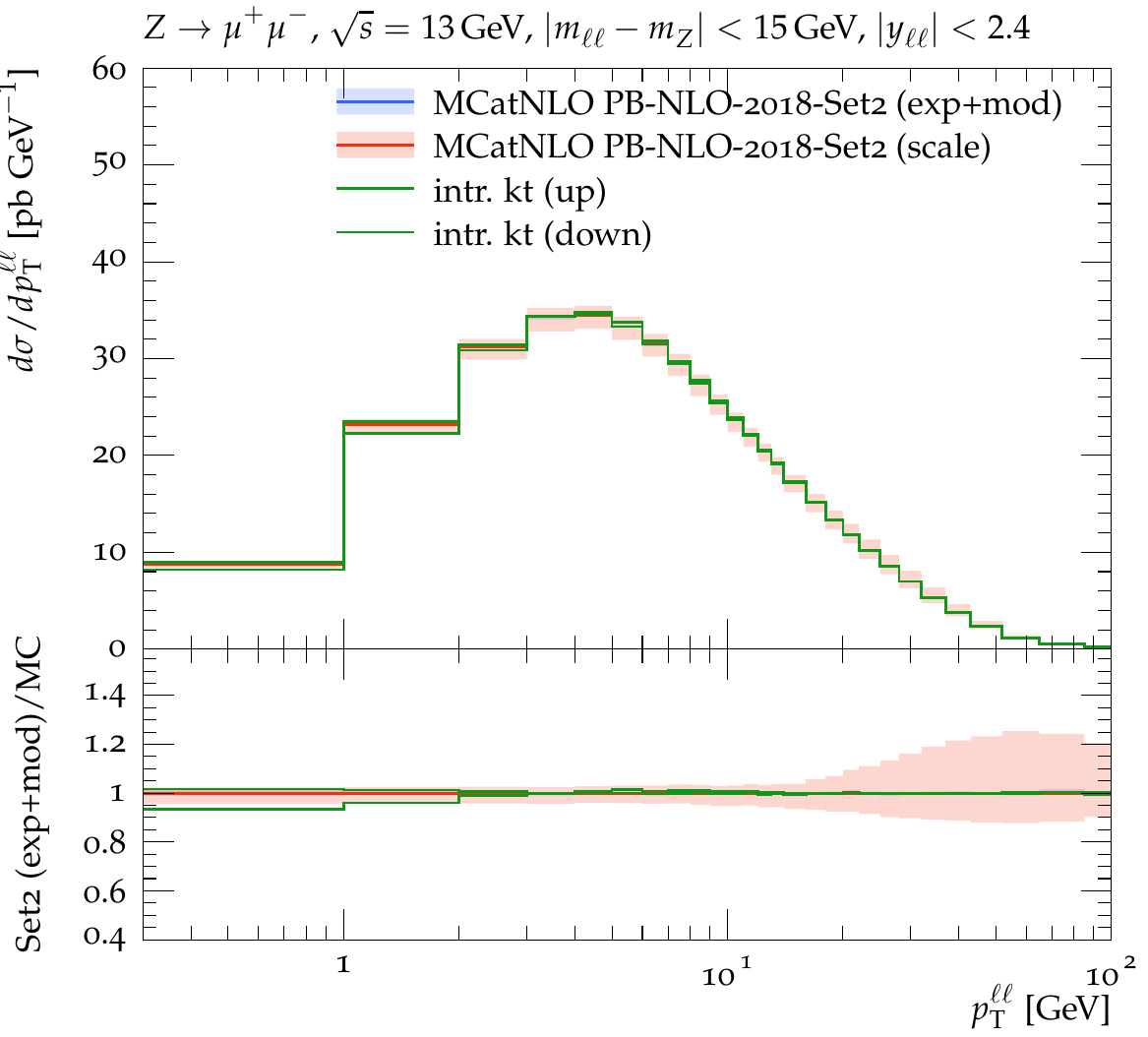}
\includegraphics[width=0.495\textwidth]{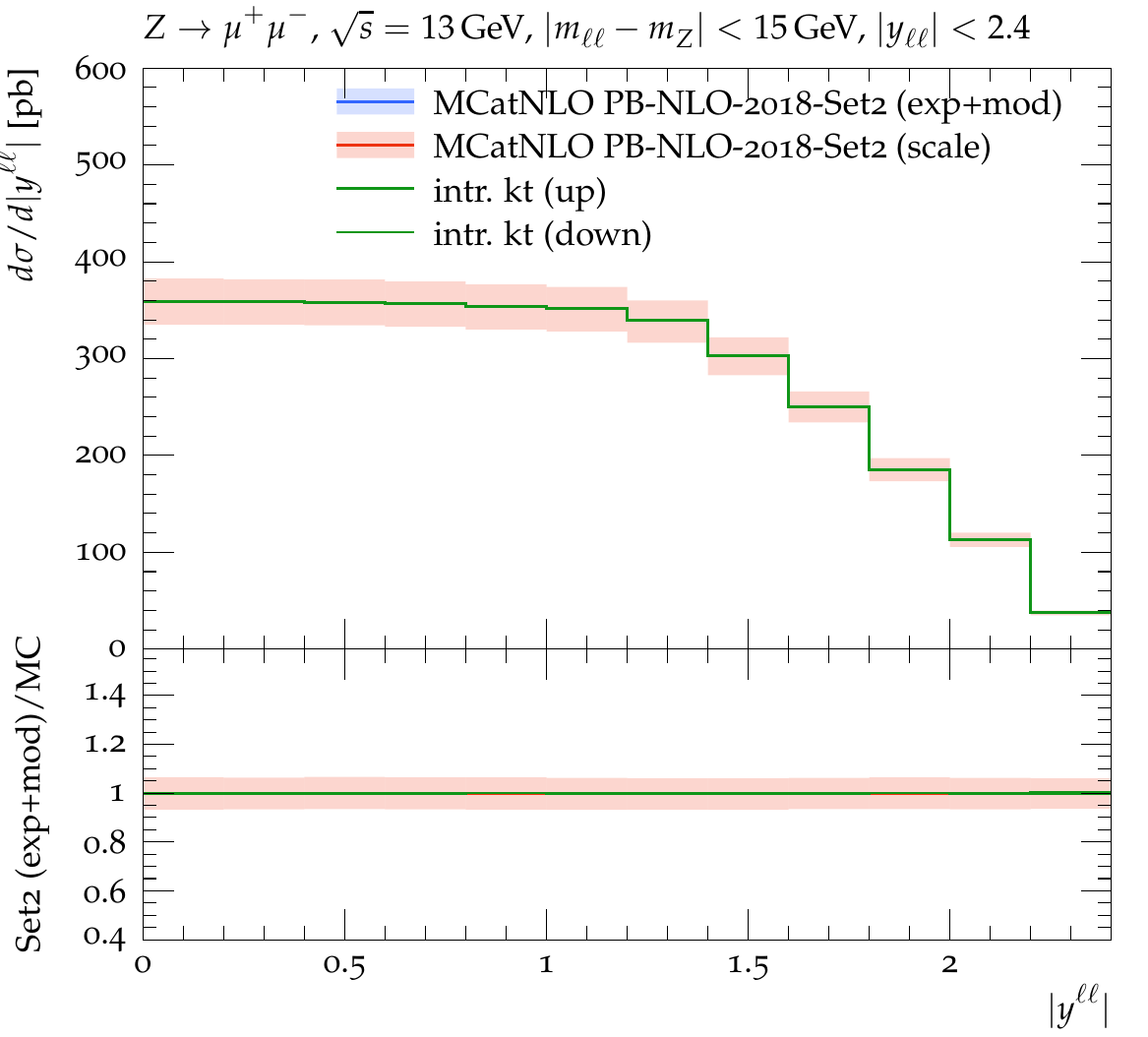}
\caption{\small Transverse momentum $p_T$ (left) and rapidity $y$ spectra of \PZ -bosons at $\sqrt{s}=13~\TeV$ from  the prediction after including TMDs.
The pdf (not visible) and the scale uncertainties are shown. In addition shown are predictions when the mean of the intrinsic gauss distribution is varied by a factor of 2 up and down.
  }
\label{Zpt-TMD_uncertainty-13TeV} 
\end{center}
\end{figure}

In Fig.~\ref{Zpt-TMD_uncertainty-13TeV} we show  predictions  for the transverse momentum and rapidity spectra of \PZ -bosons at $\sqrt{s}=13~\TeV$  obtained, as in the previous subsection,  with a calculation using \mcatnlo\ together with the \PBM -TMD~Set~2. The \PZ -bosons are selected from decay leptons with $\pt > 25~\GeV$ and $|\eta|< 2.4$ and $| m_{ll} -m_{\PZ } |<15~\GeV$, following closely the selection in \cite{CMS:2019yfw}. 
The uncertainties coming from the TMD parton density are shown as the blue band, while the uncertainties  from a variation of the factorisation and renormalization scales are shown as the red band.
In addition  are shown predictions coming from a variation of the mean of the intrinsic \kt -distribution by a factor of two up and down.

\begin{figure}[htb]
\begin{center} 
\includegraphics[width=0.495\textwidth]{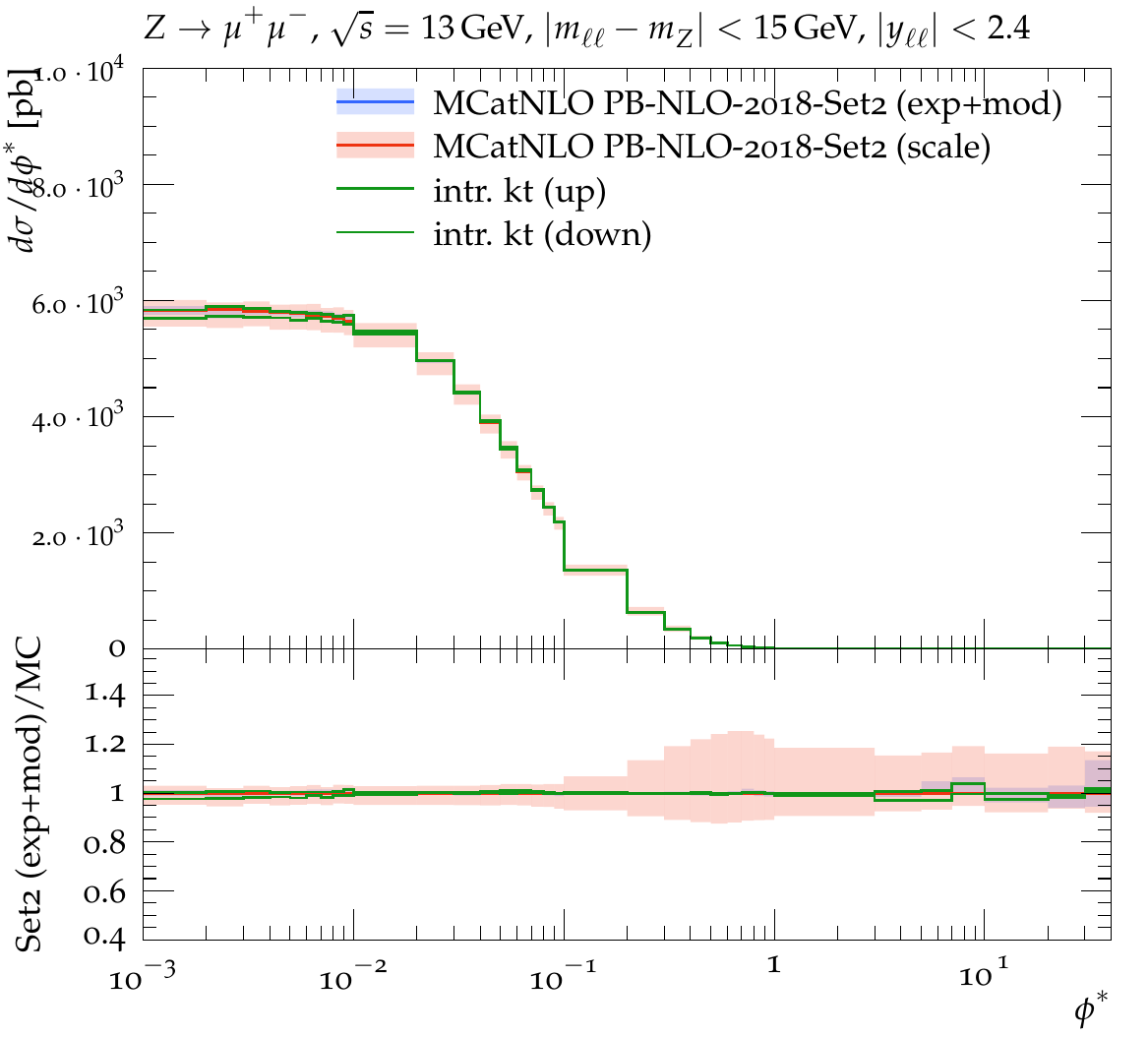}
\caption{\small  $\phi^*$ spectrum of \PZ -bosons at $\sqrt{s}=13~\TeV$ obtained from \protect\mcatnlo\ with \protect\PBM -TMD NLO 2018 Set2 \protect\cite{Martinez:2018jxt}.
The pdf (not visible) and the scale uncertainties are shown.  In addition shown are predictions when the mean of the intrinsic gauss distribution is varied by a factor of 2 up and down.
  }
\label{Zphi-TMD_uncertainty-13TeV} 
\end{center}
\end{figure}

In Fig.~\ref{Zphi-TMD_uncertainty-13TeV}  the prediction for  $\phi^*$ is shown, including TMD and scale uncertainties.

\subsection{The region of   small transverse momenta}
\begin{figure}[htb]
\begin{center} 
\includegraphics[width=0.495\textwidth]{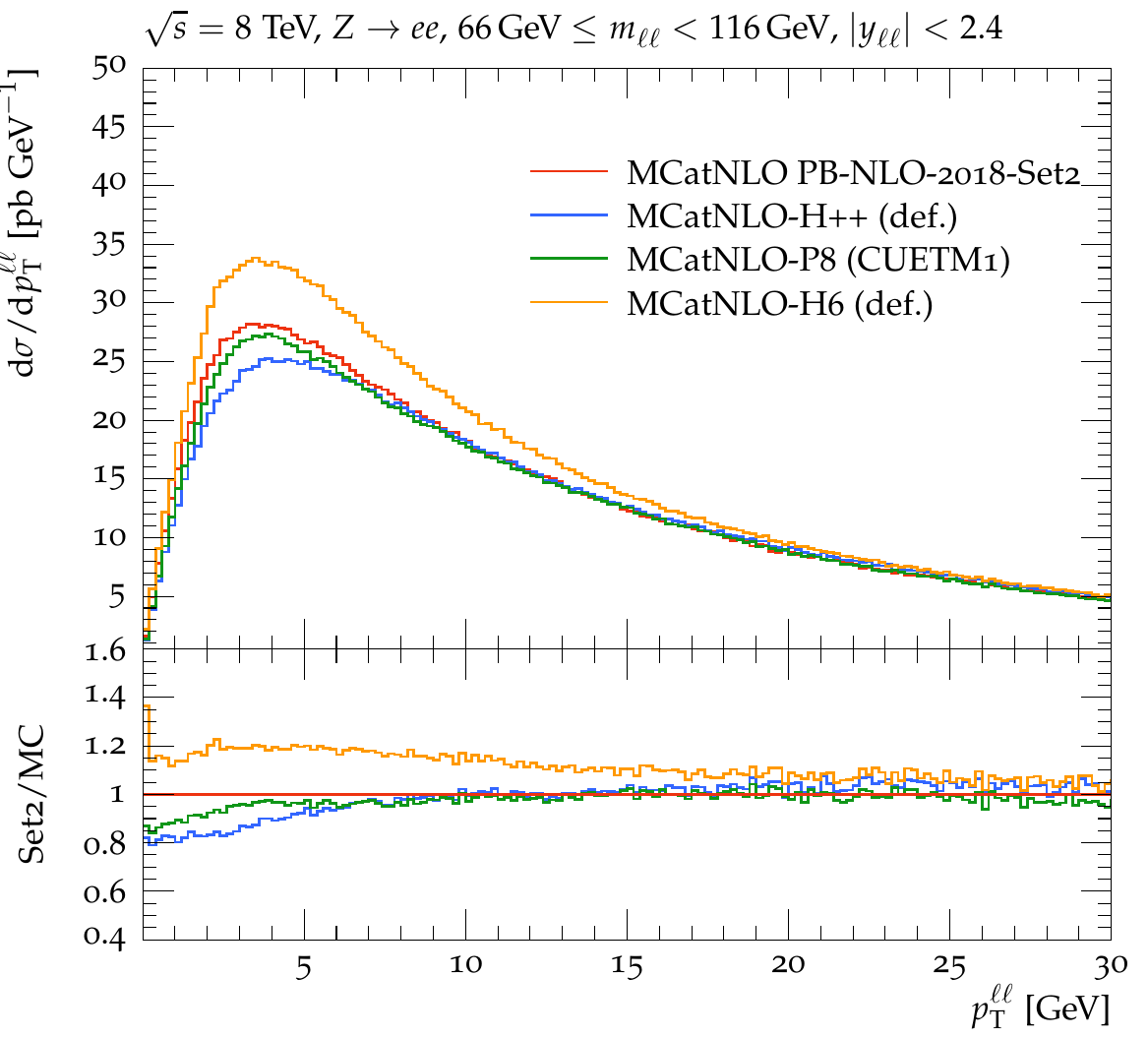} 
\includegraphics[width=0.495\textwidth]{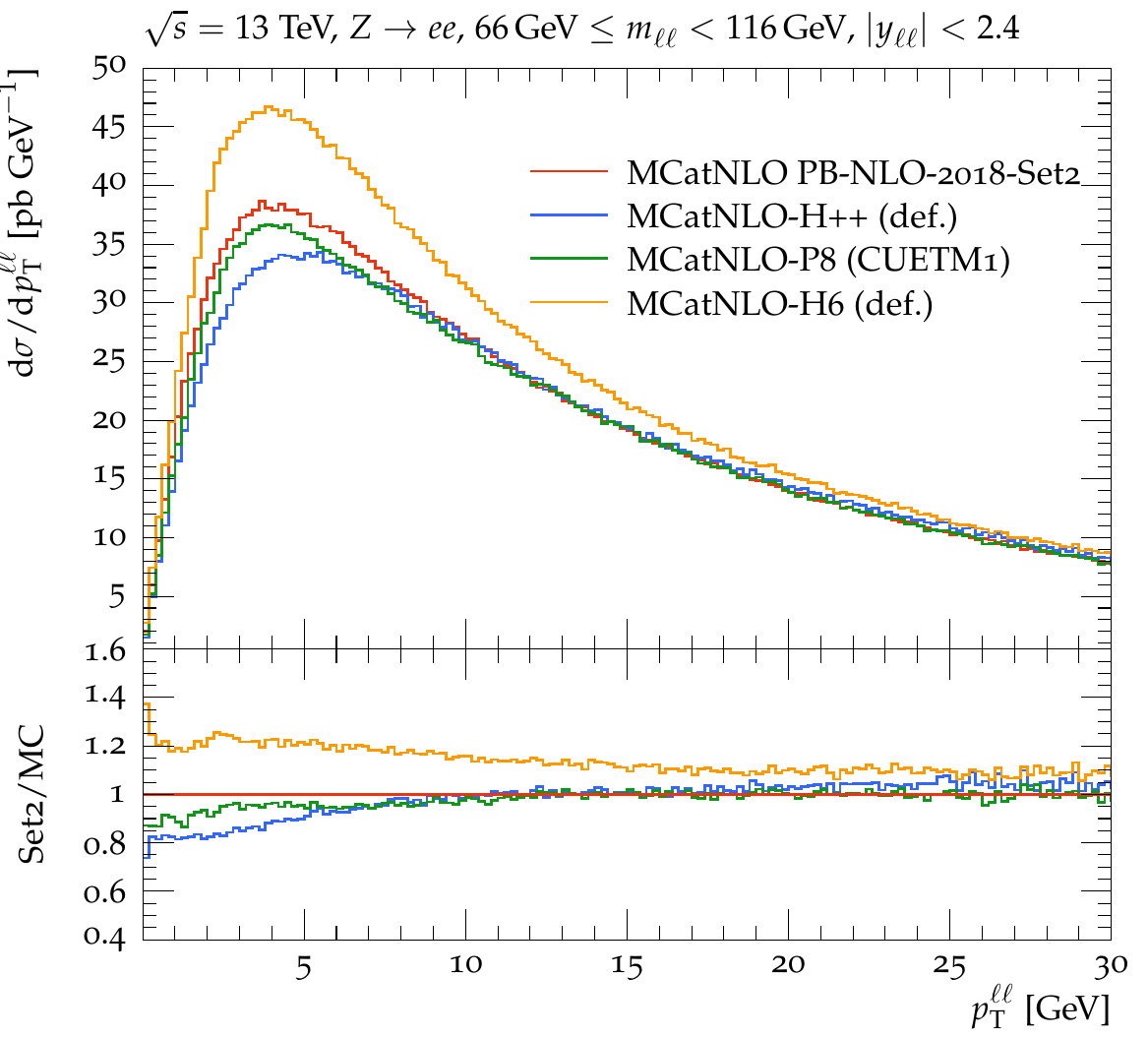} 
\caption{\small Transverse momentum $p_T^{ll}$ spectrum of \PZ -bosons at $\sqrt{s} = 8~\TeV$(left) and $13~\TeV$(right) obtained with the \PBM\ - method (\protect\PBM -TMD NLO 2018 Set2 \protect\cite{Martinez:2018jxt}), the parton shower of {\sc Pythia}8~\protect\cite{Sjostrand:2014zea} with tune CUETP8M1~\protect\cite{Khachatryan:2015pea},
{\sc Herwig}++~\protect\cite{Bahr:2008pv} , and {\sc Herwig}6~\protect\cite{Corcella:2002jc}.  }
\label{Zpt-FineBin} 
\end{center}
\end{figure}

The transverse momentum spectrum of the \PZ -bosons at small \pt\ is a direct measure of the intrinsic motion of the partons inside the protons as well as an important probe of the perturbative soft-gluon resummation, either in terms of TMDs or in terms of parton showers. In Fig.~\ref{Zpt-FineBin} we show predictions for the \pt\ spectrum with a fine binning obtained with  the \PBM\ - method and standard parton showers with recent tunes: \pythia 8~\cite{Sjostrand:2014zea} with tune CUETP8M1~\cite{Khachatryan:2015pea}, \herwig ++~\cite{Bahr:2008pv}  and \herwig 6~\cite{Corcella:2002jc} with default parameter settings (version 6.5.21 in {\sc MadGraph5\_aMC@NLO} (version 2.6.4)). All predictions make use of \mcatnlo\ calculations of the hard process  with the same collinear parton density (PB-NLO-2018-Set2) but with the appropriate subtraction terms included in the \mcatnlo\ calculation. As expected, all calculations agree at larger \pt\ , while  differences of up to 20~\% are observed at small $\pt < 5~\GeV$. The prediction using \herwig 6 uses parameter settings  which were not tuned to recent measurements, and serves as an illustration of the sensitivity of MC tunes. With dedicated measurements in the region of  \PZ -boson $\pt < 5-10~\GeV $ with fine enough binning, differences in the resummations and parton showers can be distinguished.

\section{Conclusion}

The \PBM -TMDs, which have been determined from NLO fits to inclusive DIS data, have been used to predict the \PZ -boson transverse momentum spectrum in \Pp{}\Pp\ collisions at the LHC  for $\sqrt{s}=8$ and $13~\TeV$, using NLO collinear matrix element calculations for inclusive \PZ -production. The \mcatnlo\ framework as implemented in {\sc MadGraph5\_aMC@NLO} with the \herwig 6 subtraction terms is used. The \PBM -TMDs are used to generate transverse momenta of the incoming partons. The matching of \PBM -TMDs with the NLO calculation is performed by  defining the factorisation scale $\mu$ for the TMD, different for the Born and real emission contributions.

The principle on which the PB method is based is similar to that of parton showers, but the difference is that in the PB method TMD densities are defined and determined from fits to experimental data, which places constraints on fixed-scale inputs to evolution. 

The prediction for \PZ -boson production obtained with  \PBM -TMD together with \mcatnlo\ has been compared to measurements of ATLAS at $\sqrt{s}=8~\TeV$, and very good agreement with the measurement at small \pt\ is found using the PB-NLO-2018~Set2. The uncertainties coming from the determination of the \PBM -TMDs are quite small, and only in the lowest \pt -region a sensitivity to the intrinsic transverse momentum spectrum is found, of the order of 2--3 \%. The scale dependence of the matrix element calculation dominates the overall uncertainty.

The  \PBM -TMD combined with \mcatnlo\ has been used to predict the transverse momentum spectrum of \PZ -bosons at 13~\TeV. Measurements  of the transverse momentum spectrum with a very fine binning at small \pt\ would allow one to separate details of the resummation procedure and the intrinsic transverse momentum distributions.

The \PBM -TMD have been  combined for the first time with NLO collinear matrix element calculations and very good agreement with measurements for the transverse momentum and rapidity cross sections of \PZ -boson production is observed, without further adjusting any parameters, in contrast to what is needed in traditional parton shower approaches.

\vskip 1 cm 

\noindent 
{\bf Acknowledgments.}
We are grateful to R. Frederix for many discussions and advice on aMCatNLO, and to 
A.~Apyan and  G.~Gomez Ceballos Retuerto for  
 very useful comments on the manuscript. 
FH acknowledges the support and hospitality of  DESY, Hamburg  while part of this work was being done. 
HJU thanks the Polish Science and Humboldt Foundations for the Humboldt Research fellowship during which part of this work was completed and K. Kutak for many valuable and instructive discussions.
STM thanks the Humboldt Foundation for the Georg Forster research fellowship  and 
gratefully acknowledges support from IPM.

\vskip 0.6cm 

\providecommand{\href}[2]{#2}\begingroup\raggedright\endgroup

\end{document}